\documentstyle[aasms4,amstex,amsfonts,epsfig,rotating,float]{article}

\lefthead{Melia, Liu \& Coker}
\righthead{Polarized mm and sub-mm Emission from Sgr A*}

\begin{document}
\centerline{Submitted to the Astrophysical Journal {\it Letters}}
\bigskip
\title{Polarized mm And sub-mm Emission From Sgr A* At The Galactic Center}

\author{Fulvio Melia$^{\dag}$\altaffilmark{1}}
\affil{$^{\dag}$Physics Department and Steward Observatory, The University of 
Arizona, Tucson, AZ 85721}

\altaffiltext{1}{Sir Thomas Lyle Fellow and Miegunyah Fellow.}

\author{Siming Liu$^*$}
\affil{$^*$Physics Department, The University of Arizona, Tucson, AZ 85721}

\and

\author{Robert Coker$^\ddag$}
\affil{$^\ddag$Department of Physics \& Astronomy, The University of Leeds,
Leeds  LS2 9JT, UK}

\begin{abstract}
The recent detection of significant linear polarization at mm and
sub-mm wavelengths in the spectrum of Sgr A* (if confirmed) will be
a useful probe of the conditions within several Schwarzschild radii
($r_S$) of the event horizon at the Galactic Center.  Hydrodynamic 
simulations of gas flowing in the vicinity of this object suggest that 
the infalling gas circularizes when it approaches within $5-25\;r_S$
of the black hole. We suggest that the sub-mm ``excess'' of emission 
seen in the spectrum of Sgr A* may be associated with radiation produced 
within the inner Keplerian region and that the observed polarization 
characteristics provide direct evidence for this phenomenon. The overall
spectrum from this region, including the high-energy component due
to bremsstrahlung and inverse Compton scattering processes, is at
or below the recent {\it Chandra} measurement, and may account for
the X-ray source if it turns out to be the actual counterpart to Sgr A*. 

\end{abstract}


\section{INTRODUCTION}
Discovered over 25 years ago (Balick \& Brown 1974),
Sgr A$^*$ is a bright, compact radio source coincident with the dynamical center of
the Galaxy, and now provides possibly the most compelling evidence for the
existence of supermassive black holes. The suggested central dark mass concentration 
within the inner 0.015 pc of the Galactic center is $2.6\pm0.2\times 10^6\;M_\odot$ 
(Genzel et al. 1996; Eckart \& Genzel 1996; Eckart \& Genzel 1997; Ghez et al. 
1998).  (0.1$''$ corresponds to 800 Astronomical Units, or roughly 
$1.2\times10^{16}$ cm at a distance of 8.5 kpc.) Most of this mass is probably 
associated with Sgr A*.

The spectrum of this unusual object is seen to be bumpy, but can be described as a
power-law with a spectral index $a$ that varies between roughly $0.19-0.34$
($S_{\nu}\propto\nu^a$) between GHz and mm wavelengths. However, one of 
the most interesting features currently under focus is the suggestion of a 
sub-millimeter (sub-mm) bump in the spectrum (Zylka et al. 1992; Zylka et al. 
1995), since in all emission models the highest frequencies correspond to the 
smallest spatial scales, so that the sub-millimeter
emission comes directly from the vicinity of the black hole
(Melia 1992, 1994; Melia, Jokipii \& Narayanan 1992; Coker and Melia 2000). 
The existence of this bump (or ``excess'')
has been uncertain due to the variability of Sgr A*, but is now well established 
following a set of simultaneous observations (from $\lambda$20cm to $\lambda$1mm) 
using the VLA, BIMA, Nobeyama 45 m, \& IRAM 30 m telescopes (Falcke, et al. 1998).

More recently, radio observations of Sgr A* have focused on the detection
of polarization from this source.  Although the upper
limits to the linear polarization in Sgr A* are found to be quite low (less than $1\%$)
below 86 GHz (Bower et al. 1999), this is not the case at
750, 850, 1350, and 2000 $\mu$m, where a surprisingly large intrinsic
polarization of over $10\%$ has now been reported (Aitken, et al. 2000).
>From the lack of polarization at longer wavelengths, Aitken et al. conclude
that their measured values at higher frequencies must arise in the mm/sub-mm ``excess''.
These observations also point to the tantalizing result that the position
angle changes considerably (by about $80^o$) between the mm and the sub-mm
portions of the spectrum, which one would think must surely have something to do 
with the fact that the emitting gas becomes transparent at sub-mm wavelengths
(Melia 1992, 1994).

In a companion paper (Melia, Liu \& Coker 2001) we suggested that the mm
and sub-mm ``excess'' in the spectrum of Sgr A* may be the first indirect evidence 
for the anticipated circularization of the gas falling into the black hole at 
$5-25\;r_S$, where $r_S\equiv 2GM/c^2$ is the Schwarzschild radius. 
The abundance of gas in the environment surrounding Sgr A* clearly points to accretion
as the incipient cause of the ensuing energetic behavior of this
source (Melia 1994)---whether or not it eventually leads to expulsion of some 
plasma at smaller radii (see, e.g., Falcke et al. 1993).  In their 
simulation of the Bondi-Hoyle
accretion onto Sgr A* from the surrounding winds, Coker \& Melia (1997) concluded
the accreted specific angular momentum $l\equiv \lambda r_S c$ can 
vary by $50\%$ over ${\lower.5ex\hbox{$\; \buildrel < \over \sim \;$}}$
200 years with an average equilibrium value in $\lambda$ of about $30$ or less.
Although this shows that relatively little specific angular momentum is accreted---so 
that large disks (such as those required in ADAF models; Narayan et al. 1996) probably
do not form around Sgr A*---it does nonetheless lead to the expectation that
the plasma must circularize toward smaller radii before flowing through the
event horizon.  However, given the fluctuations in the accreted value of
$\lambda$ (both in magnitude and sign!), this Keplerian flow is variable, and it
probably dissolves and reforms (possibly with a different sense of spin)
on a time scale of $\sim 100$ years or less.

Melia et al. (2001) demonstrated how this dichotomy comprising a quasi-spherical
flow at radii beyond $50\; r_S$ or so, and a Keplerian structure toward smaller
radii, may be the explanation for Sgr A*'s spectrum, including the appearance of
the ``excess'', which is viewed as arising primarily within the circularized component. 
It is our intention in this {\it Letter} to demonstrate how the linear polarization data 
may now be taken as direct evidence for the existence of this accretion profile.

\section{METHODOLOGY}

The structure of the flow within the circularization radius (at
$\sim 5-50\;r_S$) is developed fully in Melia et al. (2001).  For the
sake of completeness, we highlight the key elements of this calculation below.
Central to the modeling of the sub-mm ``excess'' is the supposition that
within the Keplerian flow, a magnetohydrodynamic dynamo produces an enhanced 
(though still sub-equipartition) magnetic field, dominated by its azimuthal 
component (Hawley, Gammie \& Balbus 1996).  Although the process of magnetic field 
dissipation suppresses the field intensity well below its equipartition value
in the quasi-spherical region (Kowalenko \& Melia 2000), the magnetic
dynamo evidently overwhelms the rate of field destruction in the differentially rotating
portion of the inflow, and the field reaches a saturated intensity.  The
sub-mm ``excess'' in Sgr A* may be the thermal synchrotron radiation produced
in this inner region. 

At a (cylindrical) radius $r$ in the Keplerian flow where the column density is 
$\Sigma$ and the angular velocity is $\Omega=(GM/r^3)^{1/2}$, the radial 
velocity is given as (e.g., Stoeger 1980) 
\begin{equation}
v_r=-{3\over r^{1/2}\Sigma}\,{\partial\over\partial r}\left(\nu\Sigma 
r^{1/2}\right)\;,\label{vr}
\end{equation}
where $\nu=(2/3)W_{r\phi}/\Sigma\;\Omega$
is the kinematic viscosity, and $W_{r\phi}$ is the vertically integrated sum of 
the Maxwell and Reynolds stresses (Balbus et al. 1994).  For the problem at hand,
the Maxwell stress dominates, and
\begin{equation}
W_{r\phi}\approx\beta_\nu\int dz\;\langle{B^2\over 8\pi}\rangle\label{visc}\;,
\end{equation}
where $B$ is the turbulent magnetic field (the average inside the integral 
being taken over time).  Numerical simulations 
(e.g., by Brandenburg et al. 1995) show that $\beta_\nu$ changes very slowly with 
$r$.  In the particular cases considered by these authors, $\beta_\nu$ ranged 
in value from $\approx 0.1$ to $0.2$, while $r$ decreased by a factor of $5$.  
For simplicity, we will here adopt a ``mean'' value of $\sim 0.15$ for this quantity.

For steady conditions, one can obtain the vertical profile by
assuming that the gas is in local hydrostatic equilibrium. Balancing
gravity and the pressure gradient in the vertical direction,
we obtain the scale height $H=\sqrt{2 R_g T r^3/\mu GM}$,
where $T$ is the gas temperature at radius r, $R_g$ is the gas constant, and $\mu$ is the
molecular weight.  For simplicity, we will assume 
that the Keplerian flow is axisymmetric and is independent of the vertical 
coordinate. Written another way, we have $(H\Omega)^2=2P/\rho$, where $P$ is the gas 
pressure and $\rho$ is the mass density of the gas. The numerical
simulations of the magnetohydrodynamic dynamo effect indicate that the field
intensity is somewhat below its equipartition value, so that
\begin{equation}
\int dz\;\langle{B^2\over 8\pi}\rangle\approx
\beta_p\int P\;dz=\beta_p{R_g\Sigma T\over\mu}\;, \label{mag}
\end{equation}
where $\beta_p$ is roughly constant with a value of $\approx 0.02$.
Thus, with $\dot M=-2\pi\,r\,\Sigma\,v_r$, we can integrate Equation (\ref{vr}) to obtain
$v_r$, and $T$ follows directly from the energy conservation
equation (Melia et al 2001).

\section{CALCULATION OF THE SPECTRUM}

The flux density (at earth) produced by the Keplerian portion of
the flow is given by
\begin{equation}
F_{\nu_0}={1\over D^2}\int I_{\nu^\prime} \sqrt{1-r_S/r}\ dA\, ,
\end{equation}
where $D=8.5$ kpc is the distance to the Galactic Center,
$\nu_0$ is the observed frequency at infinity and $\nu^\prime$ is
the frequency measured by a stationary observer in the Schwarzschild frame.
(For simplicity, we here assume the metric for a non-spinning black hole.
A more thorough exploration of the parameter values, including the black
hole spin, will be discussed elsewhere.) The frequency transformations 
are given by
\begin{equation}
\nu_0 = \nu^\prime \sqrt{1-r_S/r}\;,\qquad
\nu^\prime = \nu {\sqrt{1-v_\phi^2/c^2}\over 1-(v_\phi/c)\cos{\theta}}\;,
\end{equation}
where $\nu$ is the frequency measured in the co-moving frame, and $\theta$ is the
angle between
the velocity $\vec v_\phi$ and the line of sight. Since the radial velocity is
always much smaller than $v_\phi$, we ignore this component in the transformation
equations. So $\cos{\theta} = \sin{i}\;\cos{\phi}$,
where $i$ is the inclination angle of the axis perpendicular to
the Keplerian flow, and $\phi$ is the azimuth of the emitting element.
When the Doppler shift is included, the blue shifted region is located primarily
near $\phi=0$ while the red shifted region is at $\phi=\pi$.
The other quantities that are necessary for an evaluation of the flux
density are the area element
$dA = {\sqrt{1-r_S/r}}^{-1}\,\cos{i}\ r\ dr\ d\phi$,
and the specific intensity $I_{\nu^\prime} = 
B^\prime_{\nu^\prime}(1-e^{-\tau})$, where
\begin{equation}
B^\prime_{\nu^\prime}= \left({\sqrt{1-v_\phi^2/c^2}\over
1-(v_\phi/c)\cos{\theta}}\right)^3 B_\nu\;,
\end{equation}
and the optical depth is
\begin{equation}
\tau=\int \kappa^\prime_{\nu^\prime}\;ds = \kappa_\nu\;{2H\over
\cos{i}}\;{1-(v_\phi/c)\cos{\theta}\over \sqrt{1-v_\phi^2/c^2}}\;, \label{depth}
\end{equation}
where $\kappa_\nu$ is the absorption coefficient.  When $\tau\ll 1$, 
Kirchoff's law allows us to write
\begin{equation}
I_{\nu^\prime}\approx B^\prime_{\nu^\prime} \tau=\epsilon_\nu{2H\over
\cos{i}}\left({\sqrt{1-v_\phi^2/c^2}\over
1-(v_\phi/c)\cos{\theta}}\right)^2\;, \label{Intensity2}
\end{equation}
where $\epsilon_\nu = B_\nu\ \kappa_\nu$ is the emissivity.

The presence of a substantial azimuthal component of the magnetic field
makes it convenient to calculate the observed flux directly from the
Extraordinary and Ordinary components of the intensity.	 The most
convenient approach is to select the symmetry axis of the Keplerian
flow as the reference direction. The observed flux densities in the 
azimuthal and the reference directions are given by
\begin{eqnarray}
F_{1\nu_0}&=& {1\over D^2}\int (I^e_{\nu^\prime}\cos^2{\phi^\prime}+
I^o_{\nu^\prime}\sin^2{\phi^\prime})\sqrt{1-r_S/r}\ dA\ ,\\
F_{2\nu_0}&=& {1\over D^2}\int (I^e_{\nu^\prime}\sin^2{\phi^\prime}+
I^o_{\nu^\prime}\cos^2{\phi^\prime})\sqrt{1-r_S/r}\ dA\ , 
\end{eqnarray}
respectively, where $\phi^\prime+\pi/2$ is the position angle of the 
magnetic field vector within the emitting element that 
has an azimuth of $\phi$, so that
$\cot{\phi^\prime}=\cot{\phi}\;\cos{i}$. $I^e_{\nu^\prime}$ and $I^o_{\nu^\prime}$ are the
specific intensities for
the Extraordinary and Ordinary waves, respectively. For thermal synchrotron 
radiation, the emissivities are 
\begin{eqnarray}
\epsilon^e&=& {\sqrt{3} e^3\over 8\pi m_e c^2} B \sin{\theta^\prime} \int_0^\infty
N(E)[F(x)+G(x)]\ dE\ , \label{com1} \\
\epsilon^o&=& {\sqrt{3} e^3\over 8\pi m_e c^2} B \sin{\theta^\prime} \int_0^\infty
N(E)[F(x)-G(x)]\ dE\;, \label{com2}
\end{eqnarray}
where $N(E)$ is the electron distribution function at energy $E$, and
\begin{eqnarray}
\cos{\theta^\prime}= {\cos{\theta}-v_\phi/c\over 1-(v_\phi/c)\cos{\theta}}\; ,
&\qquad& x={4\pi\nu m_e^3c^5\over 3eB\sin{\theta^\prime}E^2}\; ,\\
F(x)= x\int_x^\infty K_{5/3}(z)\ dz\;,&\qquad& 
G(x)= x\ K_{2/3}(x)\;.
\end{eqnarray}
$K_{5/3}$ and $K_{2/3}$ are the corresponding modified Bessel functions
(Pacholczyk 1970).  The total flux density produced by the Keplerian portion of the flow is 
the sum of these two.  The expected fractional polarization is then given by
$P_{\nu_0}=(F_{1\nu_0}-F_{2\nu_0})/(F_{1\nu_0}+F_{2\nu_0})$.

The temperature in this region reaches $\sim 10^{11}$ K, for which inverse
Compton processes must be taken into account.  The self-Comptonization
of the sub-mm radiation is calculated according to the prescription
in Melia et al. (2001), based on the algorithm described in Melia
\& Fatuzzo (1989).

\section{RESULTS AND CONCLUSIONS}

The best-fit model for the polarized mm and sub-mm emission from Sgr A* 
(Aitken et al. 2000) is shown in Figures 1 (the inset)
and (the solid curve of) 2. The peak frequency of 
the flux density is $2.4\times 10^{11}$ Hz, and the flip frequency 
(at which the position angle changes by $90^o$) is $2.8\times 10^{11}$ Hz. 
Below this frequency, the first component is smaller than the second, and
the corresponding percentage polarization is therefore (by definition)
negative.  Above the flip frequency, the first component is larger. 
Although the fit is not optimized, both the spectrum and the percentage 
polarization appear to be consistent with the data. It is to be noted that
the peak frequency is actually {\it smaller} than the flip frequency, 
which is distinct from other models that may also produce a rotation of
the position angle (see Aitken et al. 2000).

It is rather straightforward to understand the polarization characteristics
in this model. In the optically thick region (below about $1.6\times
10^{11}$ Hz), the specific intensity of the Extraordinary and Ordinary 
waves is almost isotropic in the co-moving frame because the optical
depth $\tau$ is very large. Even with the inclusion of the Doppler effect, the
emissivity of the source is relatively independent of position angle. 
But the optical depths are different for the two waves, as indicated by
Equation (\ref{depth}), and the specific intensity of the Extraordinary 
wave is slightly larger than that of the Ordinary wave. From Equations 
(\ref{com1}) and (\ref{com2}), we see that the second component is 
larger than the first, and the percentage polarization is therefore
negative according to the definition of $P_{\nu_0}$. With an increase in 
frequency, the Extraordinary amplitude becomes even larger (relative to that
of the Ordinary wave) and so the percentage polarization increases.

However, in the optically thin region, the specific intensity is given 
by Equation (\ref{Intensity2}). The synchrotron emissivity is very sensitive 
to the angle between the line of sight and the magnetic field vector {\bf B}; 
synchrotron radiation is beamed into a plane perpendicular to {\bf B}
in the co-moving frame. With the inclusion of the Doppler effect, the radiation 
is beamed into a cone, and the dominant contribution comes from the blue 
shifted region which has an azimuth of about zero. 
Therefore, since the Extraordinary wave is more intense than the Ordinary wave and the
integrals (\ref{com1}) and (\ref{com2}) are dominated by radiation from the emitting
element with an azimuth of about zero, the first component is larger than the second. 
In this case, the fractional polarization becomes positive.  

In other words, the optically thick emission is dominated by emitting elements
on the near and far sides of the black hole, for which the Extraordinary wave has a
polarization direction parallel to the reference axis. In contrast, the
dominant contribution in the thin region comes from the blue shifted
emitter to the side of the black hole, where the Extraordinary wave
has a polarization direction mostly perpendicular to this axis.
The sharp decrease in polarization at still higher frequencies is due to the
diluting effects of Comptonization emission which begins to dominate
over Synchrotron emission at that point. The inclination angle dependence of 
the fractional polarization associated with emission by the Keplerian portion
of the inflow is shown in Figure 2.  

Several issues remain to be investigated.  An important result
of our analysis is that only modest accretion rates appear to be consistent
with the polarization characteristics of Sgr A* at mm and sub-mm wavelengths.
The emitting region is compact---evidently no larger than a handful of
Schwarzschild radii.  Yet hydrodynamical simulations (Coker \& Melia 1997)
suggest that the rate at which plasma is captured at larger radii (of
order $10^4\;r_S$ or so) is several orders of magnitude higher.  If
our modeling is correct, this would seem to suggest that $\dot M$ is
variable, perhaps due to a gradual loss of mass with decreasing radius
(see, e.g., Blandford \& Begelman 1999). It is essential to self-consistently 
match the conditions within the Keplerian region of the flow with the 
quasi-spherical infall further out.  These calculations are currently 
under way, and the results will be reported elsewhere. In addition, if
the {\it Chandra} source is indeed the counterpart to Sgr A* (Baganoff et al 
2000), then the spectrum shown in Figure 1 suggests a correlated variability 
between the sub-mm and X-ray fluxes, which can be tested with the next round
of coordinated observations.

\section{ACKNOWLEDGMENTS}
We are very grateful to Marco Fatuzzo for helpful discussions.
This work was supported by a Sir Thomas Lyle Fellowship and a Miegunyah Fellowship
for distinguished overseas visitors at the University of Melbourne, and by 
NASA grants NAG5-8239 and NAG5-9205.

\clearpage
\begin{figure}[p]
\centering
{\begin{turn}{-0}
\psfig{figure=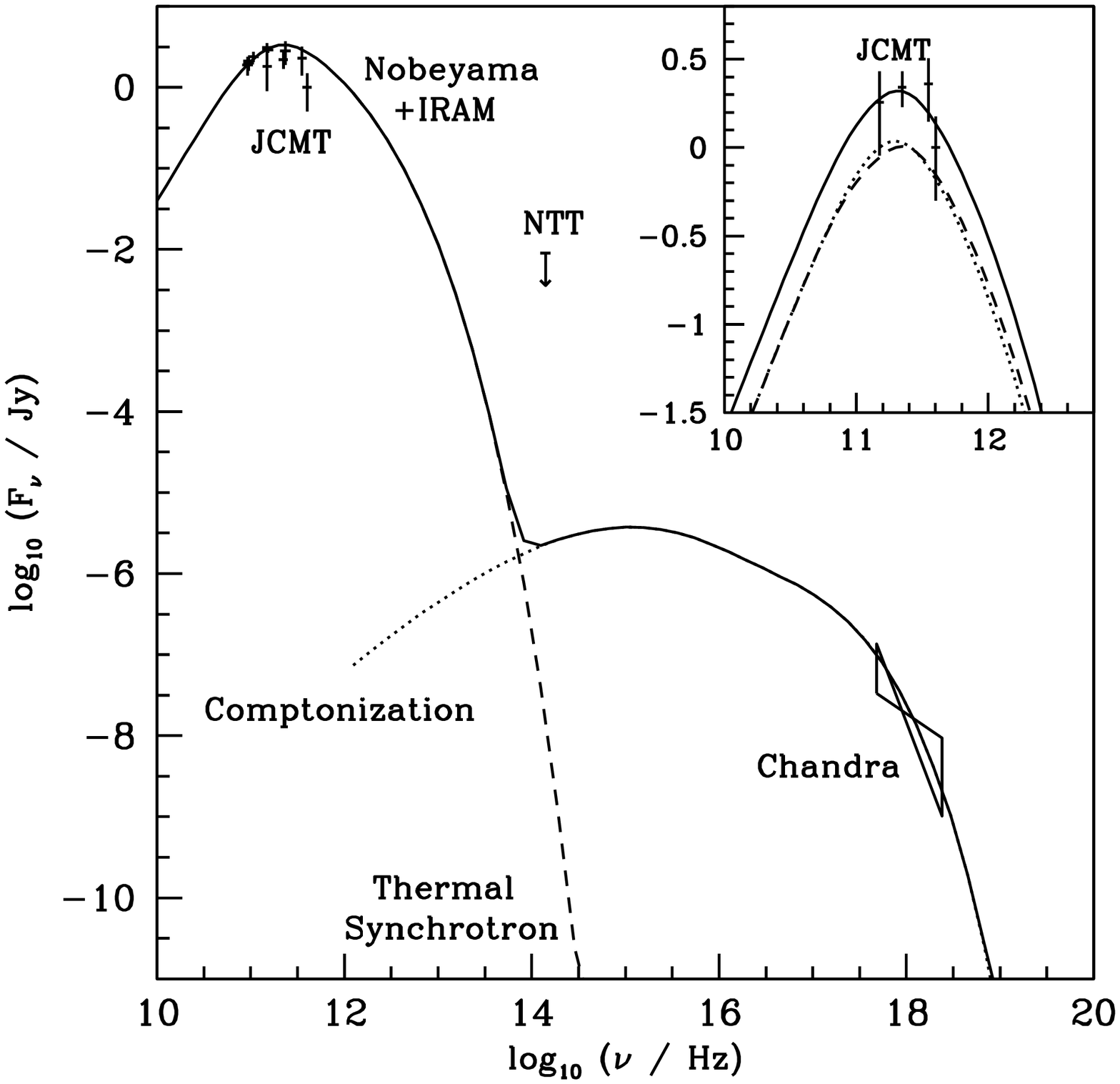,width=5.in}
\end{turn}}
\caption{The overall spectrum from this region (optimized to fit the Nobeyama
and IRAM data; Falcke et al. 1998) is shown in the main portion of this figure. 
Dashed: thermal synchrotron, dotted: self-Comptonization, solid: total spectrum.
The bremsstrahlung emission is negligible on this scale.  The parameter values 
are $\dot M=1\times 10^{16}$ g s$^{-1}$, $\beta_p=0.036$, $\beta_\nu=0.27$, 
$r_i=1.0\ r_S$ and $r_0=5.0\ r_S$.  The inclination angle of the axis perpendicular 
to the Keplerian plane is $i=60^\circ$. It is also necessary to specify the ratio of 
$v_r$ to its free-fall value at $r_0$.  For this model, this ratio is $5.0\times 10^{-5}$.
The mm to sub-mm spectrum corresponding to the best fit model for the JCMT
data (which were obtained 3 years after the Nobeyama plus IRAM observations)
is shown in the inset. The dotted curve corresponds to the first 
component and the dashed curve corresponds to the second component. The solid curve 
is the sum of these two. Here, $\dot M=4.0\times 10^{15}$ g s$^{-1}$, $\beta_p=0.02$, 
$\beta_\nu=0.15$, $r_i=1.0\ r_S$, $r_0=4.0\ r_S$ and $i=45^\circ$. The ratio of
velocities at $r_0$ is $4.0\times 10^{-5}$.}\label{fig:specbest}
\end{figure}

\begin{figure}[p]
\centering
{\begin{turn}{-0}
\psfig{figure=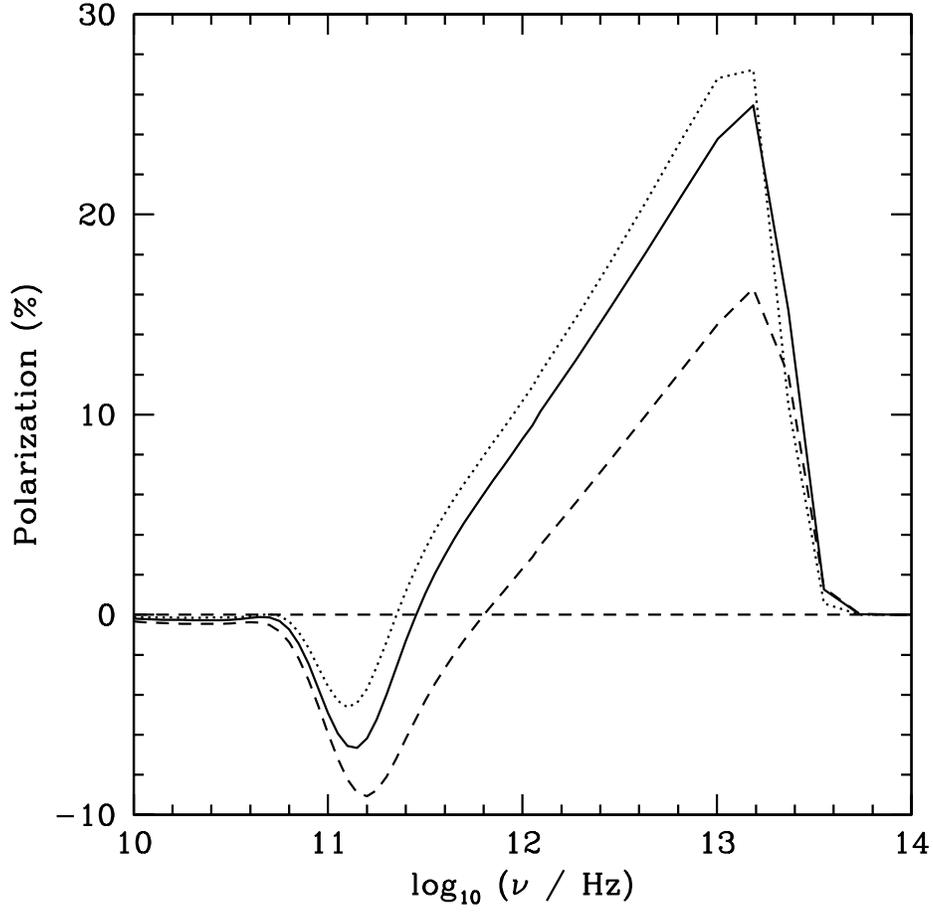,width=5.in}
\end{turn}}
\caption{The percentage polarization for the best fit model whose spectrum
is shown in the inset of Fig. 1 is here indicated by the solid curve ($i=45^\circ$).
The inclination angle dependence of this quantity is illustrated 
by the curves corresponding to an inclination angle $i=35^\circ$ (dotted),
and $i=55^\circ$ (dashed).}
\label{fig:angdependp}
\end{figure}

\end{document}